\documentclass[letterpaper, 10 pt, conference]{ieeeconf}
\IEEEoverridecommandlockouts
\usepackage{amsmath} % assumes amsmath package installed
\usepackage{amssymb}  % assumes amsmath package installed

\usepackage{amsthm}
\usepackage{mathtools}
\usepackage{verbatim} % for comments
\usepackage[linesnumbered,ruled]{algorithm2e}
\usepackage{bm}
\usepackage{graphicx,float,wrapfig,subfigure}
\usepackage{color}
\usepackage{xcolor}
\usepackage{tikz}
\usetikzlibrary{shapes.geometric, arrows}
\usepackage{epsfig} % for postscript graphics

\newtheorem{theorem}{Theorem}[section]

\usepackage{soul,color}
\usepackage{flushend}

\newcommand{\softMax}{\text{sm}}
\newcommand{\logSumExp}{\text{lse}}

\newcommand{\bydef}{\overset{\Delta}{=}}

\newcommand{\argmin}{\text{argmin}}
\newcommand{\argmax}{\text{argmax}}

\newcommand{\vecOne}{u}
\newcommand{\vecTwo}{v}
\theoremstyle{definition} %define a counter, definition
\newtheorem{definition}{Definition}[section]

\begin{document}

\title{Inducing Human Behavior to Alleviate Overstay at PEV Charging Station}% with Block Coordinate Descent}
\author{Sangjae~Bae, Teng~Zeng, Bertrand Travacca, Scott Moura% <-this % stops a space
\thanks{S. Bae, T. Zeng, B. Travacca, and S. Moura are with the Department of Civil and Environmental Engineering, University of California, Berkeley, CA, 94720 USA. \tt\{sangjae.bae, tengzeng, bertrand.travacca, smoura\}@berkeley.edu}
\thanks{This work was funded, in part, by Total S.A. and the Tsinghua-Berkeley Shenzhen Institute (TBSI).}
}

% \markboth{DRAFT. DO NOT DISTRIBUTE.}%
% {Bae \MakeLowercase{\textit{et al.}}: Inducing Human Behavior to Alleviate Overstay at PEV Charging Station}

\maketitle
\begin{abstract}
As the plug-in electric vehicle (PEV) market expands worldwide, PEV penetration has out-paced public PEV charging accessibility. In addition to charging infrastructure deployment, charging station operation is another key factor for improving charging service accessibility. In this paper, we propose a mathematical framework to optimally operate a PEV charging station, whose service capability is constrained by the number of available chargers. This mathematical framework specifically exploits human behavioral modeling to alleviate the ``overstaying'' issue that occurs when a vehicle is fully charged. Our behavioral model effectively captures human decision-making when humans are exposed to multiple charging product options, which differ in both price and quality-of-service. We reformulate the associated non-convex problem to a multi-convex problem via the Young-Fenchel transform. 
We then apply the Block Coordinate Descent algorithm to efficiently solve the optimization problem. Numerical experiments illustrate the performance of the proposed method. Simulation results show that a station operator who leverages optimally priced charging options could realize benefits in three ways: (i) net profits gains, (ii) overstay reduction, and (iii) increased quality-of-service.

% \begin{IEEEkeywords}
% plug-in electric vehicle, charging station operation, overstay, human actuated system, block coordinate descent,
% \end{IEEEkeywords}

% Two operation controllers are proposed to the respective options, namely charging with flexibility and charging as soon as possible (asap). 

% with a limited number of charging poles, which exploits behavioral analysis. A limited accessibility of charging infrastructure is one major concern to current PEV drivers. \hl{[SB: ...]} We propose a...
\end{abstract}

% \vspace{-6mm}
\section{Introduction}
% Main challenges in solving overstay problem
This paper incorporates human behavior models to address an important EV charging station operations problem -- the ``overstay" issue \cite{zeng2019solving}. This paper's goal is to propose a mathematical framework to optimally operate a charging station that maximizes operating revenue, and manages overstay duration. This optimized operation strategy may help station operators realize maximum facility utilization and customer satisfaction. Furthermore, this paper proposes an exact reformulation which turns a non-convex optimization problem into a multi-convex problem. The problem is then solved efficiently through the Block Coordinate Descent algorithm \cite{xu2013block}. This provides convergence guarantees for real-time operation.

\vspace{-2mm}
\subsection{Background \& Challenges}
A recent study based on public charging station data forecasts that the anticipated number of PEVs will reach 1 million in the U.S market by 2020, and more than 50\% of new cars sold globally will be electrified by 2040 \cite{ChargePointReport}. However, the continued growth of PEVs might be impeded by limited accessibility to charging infrastructure. Although governments and private companies have put forth great efforts to deploy public charging systems, there remains a large gap between the current service capability and the expected PEV deployment. That is, PEV penetration has out-paced charging station deployment \cite{chinaSpeedupEV,charginglocations}. In urban areas, especially central business districts, the competition for charging resources is intense. After a charger is plugged-in to a vehicle, the charger can be occupied (even if the PEV is not charging) until the driver returns from work, shopping, dining, etc. We refer to this behavior as ``overstay''. Today, ``overstay'' can consume a charger 6-8 hours in a typical day, which blocks the charger from providing charging service to other vehicles. In response to this overstay problem, the charging service providers often (i) install more chargers to satisfy demand \cite{zeng2019solving}, (ii) hire a human valet to rotate vehicles, and/or (iii) apply a steep penalty if drivers overstay.

\vspace{-2mm}
\subsection{Literature Review}
Surprisingly, very few publications have addressed the overstay issue at charging stations, despite the problem's significance. Given a constrained number of chargers at the station, one promising alternative to address accessibility is to reduce overstay duration. The state-of-art approach is to introduce a penalty on overstaying PEVs. Zeng et al. \cite{zeng2019solving} incorporates an ``interchange'' operation that actively unplugs fully charged PEVs. The paper associates the action with a financial burden, which essentially serves as a penalty to users. Biswas et al. \cite{biswas2016managing} introduces a penalty function that is activated once the actual charging session is finished, while the PEV remains occupying the charger. They develop a penalty acceptance probability to further determine the appropriate price setting. Recently, Tesla has implemented a similar approach to address this exact issue. They impose an ``idle fee," which is a penalty cost they apply to users, in [\$/minute], after the PEV is fully charged but still connected \cite{lambertTesla}. However, the effectiveness of a penalty remains to be studied. PEV drivers are individual decision makers who should be studied to understand their sensitivity to various pricing options. Furthermore, if human decision making around charging is well-understood, then station operators may optimally price charging service options to maximize profit yet ensure high station throughput.

% However, PEV would not receive charging service until penalty price is accepted in \cite{biswas2016managing}. This essentially deteriorates PEV charging station accessibility. Rather, reducing overstay duration while maintaining same level of Quality of Service (QoS) is the goal of a charging station operator.

% Incorporating human behavior into system operations

It is important to acknowledge the ``human-in-the-loop" dynamics that occur between the human customers, i.e. the PEV drivers, and the station operator. Essentially, to charge a vehicle, human drivers are exposed to prices for electricity, parking, overstaying, etc. The driver decides on a charging service accordingly, based on the delivered energy, and parking/overstay duration. Incorporating human behavior into system operations is not a new topic, particularly in the domain of Cyber-Physical \& Human Systems (CPHS). In CPHS, human inputs can be a crucial component, but play different roles in different contexts. Examples include humans as disturbances \cite{pentland1999modeling, arnold2013random}, humans as cooperator \cite{mukhopadhyay2015wearable}, or humans as system actuators \cite{bae2018acc, bae2019modeling}. In particular, in our previous work \cite{bae2019modeling}, we proposed a new perspective on human behavior in control systems, where the desired human behavior can be induced by a quantitative incentive that the system operator can control. We coin these systems as ``human actuated systems''. 

In a philosophically similar line of work, Bitar el al. \cite{bitar2016deadline} introduced a ``deadline differentiated pricing" scheme to customers for managing deferrable loads. The quantitative incentive is given by a lower electricity price when customers have later departure times, and hence endow more charging flexibility to the station operator. However, this effectively induces customers to occupy chargers longer and can exacerbate the overstay problem. In this work, we specifically address overstay -- a longer stay does not mean lower charging cost or higher profit. We propose a formalized mathematical framework to incorporate a human behavior model for a discrete set of choices, known as a Discrete Choice Model (DCM). The framework can be cast as a classical nonlinear optimal control problem, solved via dynamic programming (DP) or sequential quadratic programming (SQP). However, the scalability (in DP, due to Bellman's curse of dimensionality) or approximation errors (in SQP) remain hinder these classical numerical optimal control methods. Here, we re-formulate the non-convex optimal control problem into a multi-convex optimization program, and apply it to PEV charging station operations. 

% In this paper, we incorporate Discrete Choice Modeling (DCM) to model human decision making process and behavior to the corresponding options. We introduces varying pricing options to individual human driver with objective to induce desired behaviors.  
% \hl{Talk about lit review on DCM and system operation.}
% \hl{[SB: Incorporating human behaviors into system operations is not a new topic. In our previous paper [X], ... however not yet applied to practical applications. ]}

% Main contributions of this work
% \hl{[SB: The main contributions of this work are twofold.]}
% \subsection{Contributions}
% \vspace{-2mm}
\subsection{Contributions \& Paper Organization}
This paper provides the following novel contributions:
\begin{itemize}
    \item We identify and address an important yet poorly understood problem for PEV charging station -- overstay.
    \item We propose a mathematical framework with Discrete Choice Models to operate a PEV charging station with an optimal pricing policy. The framework incorporates a PEV driver's probability of selecting various charging options, and incorporates overstay, both of which are responsive to the pricing policy.
    \item We reformulate the non-convex operation problem into a multi-convex problem, which can be efficiently solved via Block Coordinate Descent (BCD).
    % \item We examine the proposed framework with real data-set for charging demand at a charging station located in San  Luis Obispo, California. We show promising results with significant improvement on overstaying issue as well as net profit gain at the charging station.
\end{itemize}

The remainder of the paper is organized as follows: Section \ref{sec:dcm} details the discrete choice model for charging service. Section \ref{sec:controller} formulates the station pricing controller, and Section \ref{sec:reformulation} reformulates it into multi-convex form. Section \ref{sec:reformulation} details the block coordinate descent algorithm. Section \ref{sec:simulation} presents the simulation results, using a real-world data set, and discusses limitations. Finally, conclusions are drawn in Section \ref{sec:conclusion}.

\begin{figure}
    \centering
    % \captionsetup{justification=centering}
    \includegraphics[width=1\columnwidth]{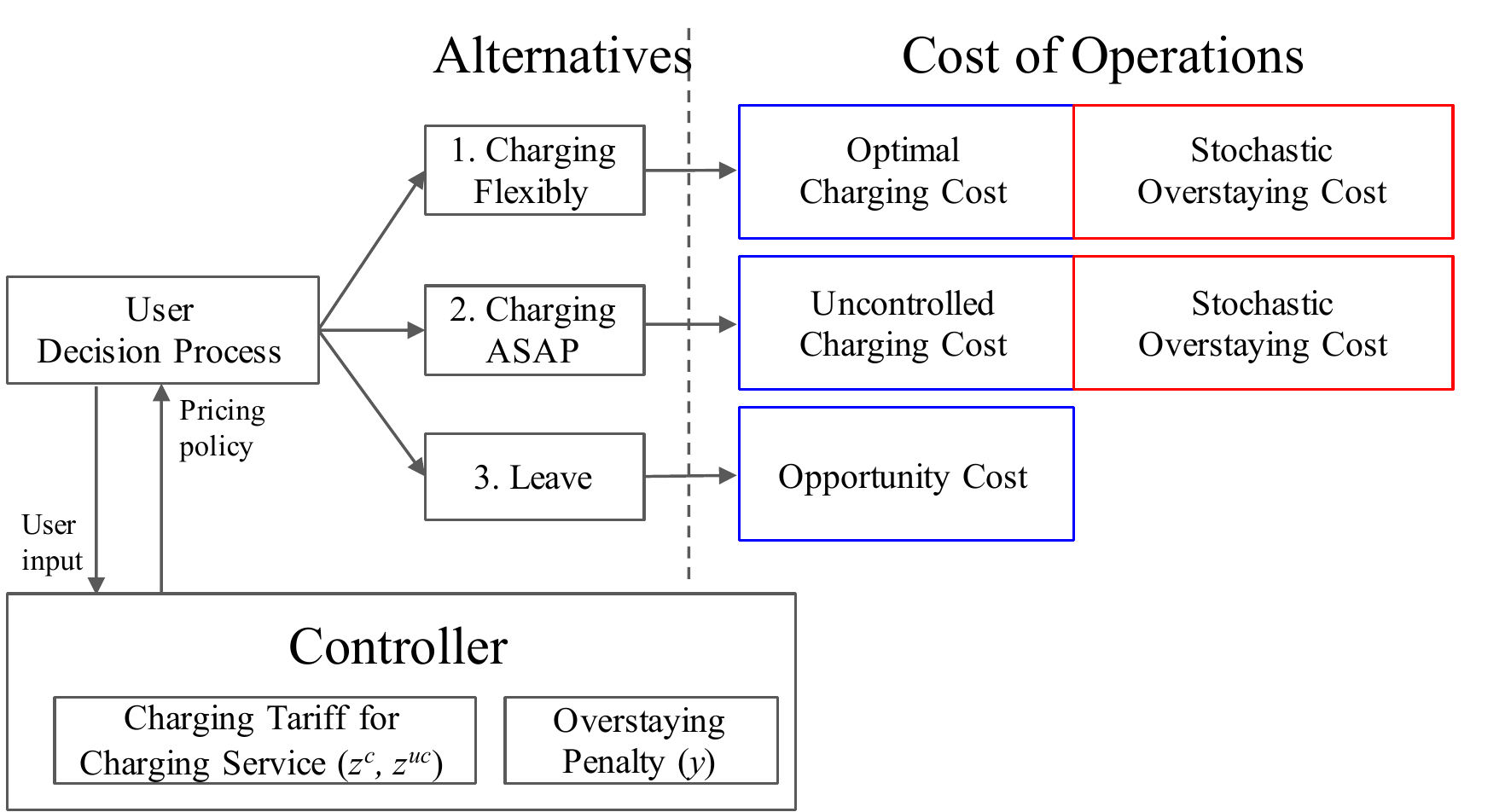}
    \caption{PEV charging station work flow: the decision process when users plug-in their PEV.}
    \label{fig:workflow}
\end{figure}

\section{Problem Overview and Behavioral Modeling}\label{sec:dcm}

\subsection{Definition and Evaluation of Overstay}
The overstay duration is defined as the time duration after a PEV is fully charged but continues to occupy a charger. In this work, we impose a penalty to overstaying PEVs. However, the overstay is evaluated differently in the two charging options (defined below), i.e., \textit{charging-flexibility} or \textit{charging-asap}. In this arrangement, a driver inputs the desired parking duration upon arrival to the charger. If the user chooses \textit{charging-flexibility}, then the overstay fee is not charged over the desired parking duration. However, if the user chooses \textit{charging-asap}, the overstay fee is charged immediately after the PEV is fully charged, irrespective of the parking duration declared by the driver upon arrival. From a station operator's perspective, incentivizing long-duration customers for flexible charging draws economic benefits by avoiding high demand charges, yet limits service to more customers.

\subsection{PEV Charging Station Operation}
Fig.~\ref{fig:workflow} illustrates the charger operation for a single PEV driver (denoted as ``user''). Upon arrival to the PEV charging station, the user inputs the following information: intended parking duration and desired added range in miles. In sequence, the user receives the pricing for two charging service options in [\$/kW], and an overstay penalty in [\$/hour] (bottom box in Fig.~\ref{fig:workflow}). These prices are computed by the pricing policy controller. Given the prices, the user chooses one of the following three options
\begin{itemize}
    \item \textit{charging-flexibility} (controlled charging, flexibility granted by customer): The needed energy is guaranteed to be delivered upon departure. However, the station operator may optimize the charging schedule.
    \item \textit{charging-asap} (uncontrolled charging, no time flexibility permitted): The PEV is charged at max power continuously, starting immediately, until the vehicle departs or the battery is full.
    \item \textit{leave}: the driver leaves the station without charging.
\end{itemize}
If a charger is vacant and a user decides for either charging service (\textit{charging-flexibility} or \textit{charging-asap}), then the charger will be occupied for the entire parking duration. When the user departs, the user pays the service fee (including overstay fees if applicable). The charger then becomes available to others. If the user decides to leave without charging (\textit{leave}), the charger remains open to others.
% For the uncontrolled charging, the overstay duration is evaluated as 

% After the PEV is fully charged, the overstay penalty is applied to 

% which will be further discussed in Section \ref{sec:controller}. That is, 

% The remaining question for the system operator is how to estimate the 

% Given those information and time-varying electricity price (i.e., Time-of-Use, TOU), the system operator determines charging tariff for two charging options: (i) charging PEV as soon as possible (asap) with the maximum charging power level (denoted as ``uncontrolled charging"); or (ii) charging PEV with flexibility during its designated parking duration (denoted as ``controlled charging"). Each charging option is associated with distinguished charging price, e.g., the user gets price discount for ``controlled charging". Given those two charging options, the user can make one of the three choices: select ``controlled charging mode'', select ``controlled charging mode'', or leave and find another charging station. Design details will be discussed in Section \ref{sec:controller}. Each action is called ``alternative'' in behavioral economics and we will use the term over the rest of the report.

\subsection{Behavioral Modeling with Discrete Choice Model (DCM)}%\label{sec:dcm}
Now, consider the perspective of the system operator. Each choice alternative is associated with a corresponding operation cost (blue and red boxes in Fig.~\ref{fig:workflow} next to each alternative). That is, evaluating which alternative the user would choose is a key problem for the system operator when determining the pricing policy. To mathematically evaluate those behaviors, we apply DCM, and more specifically the multinomial logit model \cite{train1986qualitative}. 

In DCM, each alternative has a specific utility function, and an alternative is chosen when its perceived utility is higher than that of others. Mathematically, for the $m$-th alternative, $m \in \{ 1, 2, \cdots, M \}$, the utility function is
\begin{equation}\label{eq:utility_func}
    U_m \doteq \beta_m^\top z_{m} + \gamma_m^\top w_{m} + \beta_{0m} + \epsilon_m,
\end{equation}
where $z$ is a set of ``incentive" controls, $w$ is a set of exogenous variables (or disturbances), $\beta_m$ and $\gamma_m$ are weights for the controllable inputs and uncontrollable inputs, respectively, $\beta_{0m}$ is a so-called ``alternative specific constant'', and $\epsilon_{m}$ accounts for a latent variable for unspecified errors. In the context of the PEV charging station operation, ``incentive'' controls include charging price and overstay penalty, and exogenous variables include time-of-the-day, parking duration, battery capacity, initial SOC, and needed SOC (requested by user). Interested readers can refer to \cite{bae2019modeling} for details on modeling ``incentive'' controls and exogenous variables in human actuated systems. 

Based on the multinomial logit model, the probability of choosing alternative $m$ is
\begin{align}
    \Pr(\text{alternative $m$ is chosen}) %\\
    %&\Pr \left[ \bigcap_{n \neq m} \left( U_{m} > U_{n} \right) \right] 
    = \frac{e^{V_m}}{\sum_{n=1}^M\ e^{V_n}}\label{eq:dcm_probs},
\end{align}
where $V_m \doteq \beta_m^\top z_{m} + \gamma_m^\top w_{m} + \beta_{0m}$.
Note that the probability of choosing each alternative is a sigmoid (or softmax) function, which is not convex in $z$. We will further discuss how to reformulate this non-convex function into a multi-convex optimization problem in Section \ref{sec:reformulation}.

\subsubsection{Assumptions}
We first assume that the same behavioral model is applied to all users. These users undergo the same decision process sketched in Fig.~\ref{fig:workflow} when selecting charging options, given prices. This can be relaxed by clustering users into groups. We also assume that each user faithfully chooses only one alternative at a time among the three alternatives, and that the users are rational in a way that they try to selfishly maximize their individual utilities. Those assumptions are reasonable at the PEV charging station, since it is impossible for the users to choose two charging options at the same time and they prefer the alternative that benefit themselves the most. Furthermore, we assume that the DCM parameters are known, i.e., the system operator has enough observations on the users' decisions, given different incentives, to identify an accurate DCM. Finally, we assume that the demographic information of each user is not known, i.e., only measurable data is used as features in the DCM.

% We then apply DCM, particularly multinomial logit (ML) model \cite{train1986qualitative} to mathematically evaluate behaviors of choosing charging options. 

\section{Controller Design} \label{sec:controller}
In this section, we present the detailed mathematical formulations for the pricing policy controller. Note that the control problem is solved each time a user arrives to the charging station and requests charging service.

\subsection{Control Objective}
% Main objective, high level
The objective of the pricing policy controller is to minimize a convex combination of: (i) the net charging expenses for purchasing electricity power from the grid, and (ii) the net cost associated with overstay. 
% System dynamics - states, input
We consider three control variables for the pricing policy: (i) a charging tariff for charging-flexibility $z^{\text{flex}}$, (ii) a charging tariff for charging-asap $z^{\text{asap}}$, and (iii) an overstay penalty $y$. %\hl{TZ: let's argue if the superscript $\textcolor{blue}{c}$ and $\textcolor{blue}{uc}$ need to be changed to $\textcolor{blue}{cf}$ and $\textcolor{blue}{ca}$}.

%\st{[Guys, it's confusing to have multiple sets of jargon... controlled v. uncontrolled charging, flex-charging v. charging-asap. I strongly prefer that you choose one jargon, and use only that jargon.]}
% Known Human behaviors
% We assume that the DCM parameters are known, i.e., the system operator has enough observations on the users' decisions given different charging tariff and overstay penalty, and we assume that the demographic information of each PEV driver is not known, i.e., only measurable data is used as a feature in DCM. 

% Additional constraints to consider

\subsubsection{Formulate Objective Function}
Recall that there are three possible alternatives that each user can choose: (i) charging-flexibility; (ii) charging-asap; or (iii) leave. For the first two choices, i.e., they decide to charge, the total cost is a \textit{\textcolor{blue}{deterministic cost}} resulting from provisioning electricity charging service, and a \textit{\textcolor{red}{random cost}} resulting from overstay. %\st{[Note, after you take the expectation of $T_{over}$, then $g$ is not random. It's the expected value of something random, but it's not a random variable.]} \hl{TZ: I don't fully agree. Look at Eqn.13, it's an approximation we made for simplicity. However, here is a general formulation and the indeed the overstay cost is random. I do agree we need to address this point more clear, in another way though.} 
For the third alternative, we consider a \textit{\textcolor{blue}{deterministic opportunity cost}} resulting from losing a customer. The probability of choosing each alternative is governed by the DCM, as detailed in Section \ref{sec:dcm}, which is written as a function of the control variables $z = \begin{bmatrix}z^{\text{flex}}&z^{\text{asap}}&y&1\end{bmatrix}^\top$. The objective function $J$ is the expected cost over the total parking duration, formulated as
\begin{align}
    J=&\mathbf{Pr}(\text{charging-flex})\left(\textcolor{blue}{f_{\text{flex}}}(z)+\lambda_g\textcolor{red}{g_{\text{flex}}}(z)\right) \label{eqn:obj_contolled}\\
    &+\mathbf{Pr}(\text{charging-asap})(\textcolor{blue}{f_{\text{asap}}}(z)+\lambda_g\textcolor{red}{g_{\text{asap}}}(z))\label{eqn:obj_uncontrolled}\\
    &+\mathbf{Pr}(\text{leave})(\textcolor{blue}{f_\ell}(z)), 
    \label{eqn:obj_leaving}
\end{align}
where $\textcolor{blue}{f}$ indicates a deterministic cost, $\textcolor{red}{g}$ indicates a random cost, and $\lambda_g$ is a regularization parameter. Next, we examine how the functions $f,g$ are formulated for each alternative.

\vspace{3mm}
\textit{CASE $\#1$: Selecting charging-flexibility (Eqn.\eqref{eqn:obj_contolled})}

When the user selects the charging-flexibility option, then the system operator receives added flexibility for shaping the facility's electricity demand during a designated parking duration $\Delta kN$. That is, the system operator can optimally schedule the charging profile that minimizes the electric bill cost. For the optimal charging control, we consider the following state-of-charge dynamics
\begin{equation}
    \text{SOC}_{k+1} = \text{SOC}_{k} + \frac{\Delta k \cdot \eta u_k}{B},
\end{equation}
for time step $k\in [0,N-1]$ where $u$ is the charging power level, $\Delta k$ is a time step size, $\Delta kN$ is the parking duration, $\eta \in [0,1]$ is the charger's efficiency, and $B$ is the battery capacity. 

Then $\textcolor{blue}{f_{\text{flex}}}$ is the optimal cost of charging control, formulated as
\begin{equation}
    \textcolor{blue}{f_{\text{flex}}}(z) = \min_{u,\text{SOC}} \sum_{k=0}^{N-1} u_k(c_k-z) + \lambda_u\|u\|^2, \label{eq:obj_opt_charging_linear}
    \end{equation}
    \begin{align}
    \text{subject to} \quad
    & \text{SOC}_0 = \text{SOC}_\text{init},\label{eq:const_eq_init}\\
    & \text{SOC}_{k+1} = \text{SOC}_{k} + \frac{\Delta k \eta u_k}{B}, \;\forall k\in[0,N-1], \label{eq:const_eq_dynamics}\\
    & \text{SOC}_N \geq \text{SOC}_\text{need}.\label{eq:const_ineq_SOC_need}
\end{align}
where $c_k$ is the time-of-use electric utility price at time $k$, and $\lambda_u$ is the regularization parameter for the charging power $u$. We write the equality constraints in the standard matrix form $Cx = d$, where
\begin{align*}
C &= 
\begin{bmatrix}
1&0&\cdots&0&0& 0&\cdots&0\\
-1&1&\cdots&0&0& -\frac{\Delta k\eta}{B}&\cdots&0\\
\vdots&\vdots& &\vdots&\vdots& \vdots& &\vdots\\
0&0&\cdots&-1&1& 0&\cdots&-\frac{\Delta k\eta}{B}
\end{bmatrix}\\
d &= 
\begin{bmatrix}
\text{SOC}_{\text{init}}&
0&
\cdots&
0
\end{bmatrix}^\top\\
x &= 
    \begin{bmatrix}
    \text{SOC}_0&
    \cdots&
    \text{SOC}_{N}&
    u_0&
    \cdots&
    u_{N-1}
    \end{bmatrix}^\top   
\end{align*}
The inequality constraint \eqref{eq:const_ineq_SOC_need} is also written as a matrix form $Ax \leq b$, where
\begin{equation*}
\small{
A=
    \begin{bmatrix}
    1&\cdots&0&0&0&\cdots&0\\
    \vdots& &\vdots&\vdots&\vdots& &\vdots\\
    0&\cdots&1&0&0&\cdots&0\\
    0&\cdots&0&-1&0&\cdots&0\\
     & & & & & &\\
     0&\cdots&0&0&1&\cdots&0\\
     \vdots& &\vdots&\vdots&\vdots& &\vdots\\
     0&\cdots&0&0&0&\cdots&1\\
     & & & & & &\\
     0&\cdots&0&0&-1&\cdots&0\\
     \vdots& &\vdots&\vdots& &\vdots\\
     0&\cdots&0&0&0&\cdots&-1
    \end{bmatrix}
    b=
    \begin{bmatrix}
    1\\
    % 1\\
    \vdots\\
    1\\
    -\text{SOC}_{\text{need}}\\
     \\
    p_{\max}\\
    % p_{\max}\\
    \vdots\\
    p_{\max}\\
     \\
    -p_{\min}\\
    % -p_{\min}\\
    \vdots\\
    -p_{\min}
    \end{bmatrix}}
\end{equation*}
With the dimensions $A\in \mathbb{R}^{\big((N+1)+N+N\big)\times\big((N+1)+N\big)}$ and $ B\in\mathbb{R}^{(N+1)+N+N}$. Then the optimization can be written
\begin{equation}
    f_{\text{flex}}(z) = \min_{x}\;\; \sum_{k=0}^{N-1}[x]_{N+2+k}(c_k-z^c)+\lambda_u[x]_{N+2+k}^2
\end{equation}
subject to $Ax\leq b$ and $Cx=d$. %\hl{[putting everything into matrix form is not necessary. cut if you need space.]}

Now, we evaluate the random cost due to overstay. We model the overstay duration as a Poisson process where the random variable is the overstay duration $T_{\text{overstay}}$ and the average duration $\Lambda$ is penalized by the overstay penalty $y$, e.g.,
\begin{equation}
    \Lambda(y)\coloneqq\hat{\Lambda}\frac{\hat{y}}{y}, \quad y>0,
\end{equation}
where $\hat{\Lambda}$ is an expected overstay duration without the controller, i.e., baseline, and $\hat{y}$ is a baseline overstay penalty. That is, if the optimal overstay penalty determined by the controller is equal to the baseline overstay penalty, % \hl{without control [what do you mean by without control? Do you mean the uncontrolled charging option? Again, your jargon is confusing.]}, 
the average overstay duration is the same as the baseline overstay duration. If the optimal penalty is higher than the baseline penalty, the expected overstay duration decreases in inverse proportion. The same relation is applied when the optimal penalty is lower than the baseline penalty. Then we further simplify the objective function $J$, Eqn.(\ref{eqn:obj_contolled}), by approximating the overstay cost $g_{\text{flex}}$ as the expected overstay duration, i.e.,
\begin{equation}
g_{\text{flex}} = \mathbb{E}(T_{\text{overstay}}) = \Lambda(y).  \label{eq:overstay_duration}
\end{equation}
% Evaluating the overstay duration with the Poisson process is useful in our framework since the cost \eqref{eq:overstay_duration} is convex in $z$. %We will further discuss the convexity of the problem in Section \ref{sec:reformulation}.

% Instead of using the Poisson distribution, we can utilize the empirical data from charging station and applying the logistic function (Fig.~\ref{fig:overstaying}) to estimate the overstaying duration. Within the given penalty cost of overstaying, $y$, we have an estimate of the probability. This probability will multiply with the empirical average value (or 90-percent quantile). 
% \begin{align}
%   \text{Overstaying Duration} &= P \times \Bar{T}_{\text{overstay}}\\ 
%   &= \frac{e^{5y+6}}{1+e^{5y+6}} \times \Bar{T}_{\text{overstay}},
% \end{align}
% where 5 and 6 can be tuned to get a desired shape of the probability graph. Therefore, the intuition of this probability is basically getting an estimate about the duration portion that PEV will overstay. So you could see the cost going negative from zero, as it the cost increases (the negative value decreases) the tendency to overstay loner period will drop very close to zero.

% \begin{figure}[!ht]
%     \centering
%     \includegraphics[width=.9\columnwidth]{figures/overstaying_graph.pdf}
%     \caption{Overstaying Duration Probability}
%     \label{fig:overstaying}
% \end{figure}

\vspace{3mm}
\textit{CASE $\#2$: Selecting charging-asap (Eqn.\eqref{eqn:obj_uncontrolled})}

If the user selects charging-asap, then the PEV is charged with a nominal power and the deterministic cost $f_{\text{asap}}$ is found by multiplications of parameters, i.e.,
\begin{equation}
    f_{\text{asap}}(z) =\sum_{k=1}^{ \hat{N}-1}\left(c_k-z^{\text{asap}}\right)\Delta k,
\end{equation}
where 
\begin{equation}
 \hat{N}=\frac{(\text{SOC}_{\text{need}}-\text{SOC}_{\text{init}})B}{\Delta k\eta U_{\text{nom}}}   
\end{equation}
and $U_{\text{nom}}$ is the nominal power level which is fixed and known. 
The random cost $g_{\text{asap}}$ can be formulated identically as \eqref{eq:overstay_duration} unless the system operator charges overstay differently.

\vspace{3mm}
\textit{CASE $\#3$: Leaving without charging (Eqn.\eqref{eqn:obj_leaving})}

If the user refuses to charge, then the system operator loses an opportunity to provide charging service, and consequently it creates lost revenue. We quantify this opportunity cost as the total cost of providing the uncontrolled charging without revenue,
\begin{equation}
    f_\ell(0) =\sum_{k=1}^{ \hat{N}-1}\left(c_k-0\right)\Delta k.
\end{equation}
The rationale for zero charging revenue is to quantify the opportunity cost as a fixed loss at each time step $k$, without penalizing the pricing policy $z$.

\section{Reformulation into Multi-convex Problem}\label{sec:reformulation}

Recall that the probability of choosing a given alternative \eqref{eq:dcm_probs} is not convex in $z$, and therefore the objective function \eqref{eqn:obj_contolled}-\eqref{eqn:obj_leaving} is not convex. It turns out one can reformulate these equations to yield a multi-convex problem, which can be solved via Block Coordinate Descent. This section details this process.
\vspace{-2mm}
\subsection{Objective in Compact Form}
Consider the objective function terms in \eqref{eqn:obj_contolled}- \eqref{eqn:obj_leaving} from the previous section. The objective function can be re-written in the following compact form
\begin{align}
    \min_{z\in\mathcal{Z}} & \;\; \softMax(\Theta z)_{\text{flex}} \cdot(\min_{x\in\mathcal{X}}\;h_{\text{flex}}(z,x)) \nonumber\\
    & \quad ~ +\softMax(\Theta z)_{\text{asap}} \cdot h_{\text{asap}}(z)\nonumber\\
    & \quad ~ +\softMax(\Theta z)_{\ell} \cdot h_{\ell}(z)\\
    = & \min_{z\in\mathcal{Z},x\in\mathcal{X}}\;\; \softMax(\Theta z)^\top h(z,x),\label{eq:obj_compact}
\end{align}
where $\softMax(\cdot)$ is the softmax operator
\begin{align}
\softMax(\Theta z)_{j} 
    % &= \Pr{(\theta_j^\top z=\max_{i\in \mathcal{A}} \theta_i^\top z)} \nonumber\\
    & = \frac{\exp{\theta_j^\top z}}{\sum_{i\in \mathcal{A}}\exp{\theta_i^\top z}}, \hspace{2mm} \forall j\in\mathcal{A}, \label{eqn:softmax}\\
    h(z,x)&=\begin{bmatrix}
    h_{\text{flex}}(z,x)\\
    h_{\text{asap}}(z)\\
    h_\ell(z)
    \end{bmatrix}
    =\begin{bmatrix}
    f_{\text{flex}}(x;z)+g_{\text{flex}}(z)\\
    f_{\text{asap}}(z)+g_{\text{asap}}(z)\\
    f_\ell(z)
    \end{bmatrix},\\
    z&=\begin{bmatrix}z_{\text{flex}} &z_{\text{asap}} &y &1\end{bmatrix}^\top,\\
    \Theta&=\begin{bmatrix}\theta_{\text{flex}} &\theta_{\text{asap}} &\theta_{\ell} \end{bmatrix}^\top, \mathcal{A} = \{\text{flex}, \text{asap}, \ell\},\\
    \mathcal{Z}&\;\text{is the domain of $z$}, \\
    \mathcal{X}&\;\text{is the domain of $x$, satisfying \eqref{eq:const_eq_init}-\eqref{eq:const_ineq_SOC_need}}
\end{align}
This compact notation will be useful for the subsequent derivations.

\subsection{Optimization reformulation}
To transform our non-convex problem, the following definition is useful.
\begin{definition}[\textit{Bi-convex function}]\label{def:biconvex}
    Let $X \subseteq \mathbb{R}^n$, $Y \subseteq \mathbb{R}^m$ be two non-empty, convex sets. A function $h(x,y):\mathbf{X}\times \mathbf{Y} \longrightarrow \mathbb{R}$ is called \textit{bi-convex function} if $h_x(x,y)$ is convex in $y$ for fixed $x \in \mathbf{X}$ and $h_y(x,y)$ is convex in $x$ for fixed $y \in \mathbf{Y}$. 
\end{definition}

In this work, our objective function is neither convex nor bi-convex due to the softmax function \eqref{eqn:softmax}. However, it turns out that we can exploit the problem structure and reformulate it as a \textit{multi-convex} problem, which generalizes Definition \ref{def:biconvex}. Indeed, the optimization problem can be rewritten as 

\begin{equation}
\begin{aligned}\min_{z\in\mathcal{Z},x\in\mathcal{X}}\;\; v^\top h(z,x),\\
\text{where} \hspace{2mm} v=\softMax(\Theta z), \label{eqn:original_obj}
\end{aligned}
\end{equation}
which can be converted into a 3-block \textit{multi-convex} problem. Next, we will handle the non-convex equality constraint $v=\softMax(\Theta z)$. We reformulate it as a bi-convex constraint in the next section. 

%%%%%%%%%%%%%%%%%%%%%%%%%%%%%%%%%%%%%%%%%%%%
\subsection{Bi-convex representation of the constraints}
Denote the Log-Sum-Exponential function by $\logSumExp(u) = \ln\left( \sum_{j \in \mathcal{A}} \exp(u_j) \right)$. Given $u \in \mathbb{R}^n$, we have
\begin{align*}
    \logSumExp(\vecOne) &=\ln(1^\top \exp(\vecOne)), \\
    \nabla \logSumExp(u) &= \softMax(u),\\
    \text{where} \hspace{2mm} \exp(\vecOne) &= [\exp(u_1) \dots \exp(u_n)].
\end{align*}
Also recall that by definition the convex conjugate (a.k.a. Legendre–Fenchel transformation) of Log-Sum-Exponential is 
\begin{equation}
\logSumExp^\star (\vecTwo) \bydef \max_{\vecOne} \vecOne^\top \vecTwo- \logSumExp(\vecOne)
\end{equation}
It can be shown that the conjugate of Log-Sum-Exponential is the negative entropy \cite{boyd2004convex}, i.e.,
\begin{equation}
\logSumExp^\star (\vecTwo)=\begin{cases}
\vecTwo^\top \ln(\vecTwo) \quad \text{ if } \vecTwo \geq 0 \quad \text{and } 1^\top \vecTwo=1\\
\infty \quad \quad \quad \quad ~\text{o.w.}
\end{cases}
\end{equation}
Let us define $\mathcal{V}\bydef \{v \| \ v\geq 0, \quad 1^\top v=1\}$, the set of finite discrete probability distributions. Now let us examine the Fenchel-Young inequality
\begin{align}
    \logSumExp(\vecOne)+\logSumExp^\star (\vecTwo)-\vecOne^\top \vecTwo \geq 0, \; \forall u, \; \forall v \in \mathcal{V}. \label{eqn:fenchel_young}
\end{align}
For all $v \in \mathcal{V}$, the Fenchel-Young inequality is true with equality if and only if
\begin{equation}
\vecOne= \argmax_{\vecOne} \vecOne^\top \vecTwo- \logSumExp(\vecOne),
\end{equation}
since the Log-Sum-Exponential is strictly convex. The first order optimality condition for \eqref{eqn:fenchel_young} yields
\begin{equation}
\vecTwo^*=\nabla \logSumExp(\vecOne)=\softMax(\vecOne).
\end{equation}
Hence, 
\begin{align}
    \logSumExp(\vecOne)+\logSumExp^\star (\vecTwo)-\vecOne^\top \vecTwo \leq 0 \iff \vecTwo=\softMax(\vecOne). \label{eqn:fenchel_young_inverse}
\end{align}
The significance is that we can replace the equality in (\ref{eqn:original_obj}) with the inequality constraint in \eqref{eqn:fenchel_young_inverse}. This is an equivalent reformulation, with no relaxation or approximation errors, thanks to the Fenchel-Young inequality. 

Now, replace $\vecOne$ with $\Theta z$ in (\ref{eqn:fenchel_young_inverse}):
$$v=\softMax(\Theta z) \iff \logSumExp(\Theta z)+\logSumExp^\star(v)-v^\top (\Theta z) \leq 0 $$
This inequality can be relaxed given a precision parameter $\varepsilon$ by 
$$\logSumExp(\Theta z)+\logSumExp^\star(v)-v^\top (\Theta z) \leq \varepsilon$$
This inequality defines a bi-convex set in $(z,v)$ with non-empty interior. We can then reformulate and relax the original problem into
\begin{equation}
    \begin{aligned}\min_{z\in\mathcal{Z},x\in\mathcal{X}, v\in \mathcal{V}}\;\; &v^\top h(z,x)\\
\text{subject to:} \qquad & \logSumExp(\Theta z)+\logSumExp^\star(v)-v^\top (\Theta z) \leq \varepsilon
\end{aligned}\label{eq:3_block_multiconvex}
\end{equation}
which is 3-block multiconvex in $(z,x,v)$.

\subsection{Block Coordinate Descent Algorithm}
Next, we apply the Block Coordinate Descent (BCD) algorithm \cite{xu2013block} to solve the 3-block multi-convex problem in \eqref{eq:3_block_multiconvex}. Details are illustrated in Algorithm \ref{alg:bcd}. Note that each variable update requires the solution of a convex problem. 
\begin{algorithm}
    \SetKwInOut{Input}{Input}
    \SetKwInOut{Output}{Output}
    \SetKwInOut{Init}{Init}
    \Init{$z^{(0)}=z_0, x^{(0)}=x_0, v^{(0)}=\softMax(\Theta z_0)$\\
    $F^{(0)}=v^{(0)\top} h(z^{(0)},x^{(0)})$}
    \While{$\|F^{(i+1)}-F^{(i)}\| > \epsilon$}{
    $x^{(i+1)}=\argmin_{x \in \mathcal{X}} h(z^{(i)},x)^\top v^{(i)} $ \\
    $z^{(i+1)}=\argmin_{z \in \mathcal{Z}} h(z,x^{(i+1)})^T v^{(i)}$\\ $\quad\quad\quad\quad\quad\quad\quad~~+\mu(\logSumExp(\Theta z)-z^\top \Theta^\top v^{(i)}) $\\
    $ v^{(i+1)}=\argmin_{v \in \mathcal{V}} v^\top h(z^{(i+1)},x^{(i+1)})$\\
    $\quad\quad\quad\quad\quad\quad\quad~~+\mu(\logSumExp^\star(v)-v^\top \Theta z^{(i+1)})$ \\
    }
    \caption{Block Coordinate Descent Algorithm}\label{alg:bcd}
\end{algorithm}

\subsection{BCD convergence rate}
We refer to the convergence analysis in \cite{xu2013block} and restate the analysis as follows for completeness of this paper.
\begin{theorem}
Consider an $s$-block multi-convex function $$(x_1, \cdots, x_s)\in \mathcal{X} \rightarrow f(x_1,\cdots, x_s)$$
That is the set given $j \in \{1, \cdots, s\}$ the function 
$$x_j \rightarrow f(x_1, \cdots, x_{j-1},x_j,x_{j+1}, \cdots, x_s)$$
defined on 
$$\mathcal{X}_j:=\{x_j|(x_1, \cdots, x_{j-1},x_j,x_{j+1}, \cdots, x_s)\in \mathcal{X}\}$$
is convex. If $f$ is differentiable with Lipschitz gradient, bounded below and locally strongly convex, then BCD will globally converge to a critical point of $f$ at a worst case rate $\mathcal{O}(\tau^k)$ with $0<\tau <1$ (i.e. linear convergence rate).
\end{theorem}

In our case, these hypotheses are met. Since the Log-Sum-Exponential function is locally strongly convex, the BCD algorithm must converge linearly.

\section{Simulation}\label{sec:simulation}
%To validate the effectiveness of the aforementioned control framework, we developed a agent-based simulator (in Matlab) where each user/EV driver randomly selects a charging option based on the Discrete Choice Model. Each user has unique information for charging demand. %However, recall that all the users are represented with an universal behavioral model, i.e., the sensitivity to each feature is identical among PEV drivers. 
%The personalized charging demand in our simulation is based on a real dataset from a charging station in California, and the behavioral model for the users is synthetically generated. The details are discussed next. \hl{[SB: we can remove the intro]}

\subsection{Simulation Overview}
\subsubsection{Input Data Overview}
\begin{figure}
    \centering
    \begin{subfigure}
        \centering
        \includegraphics[width=1\columnwidth]{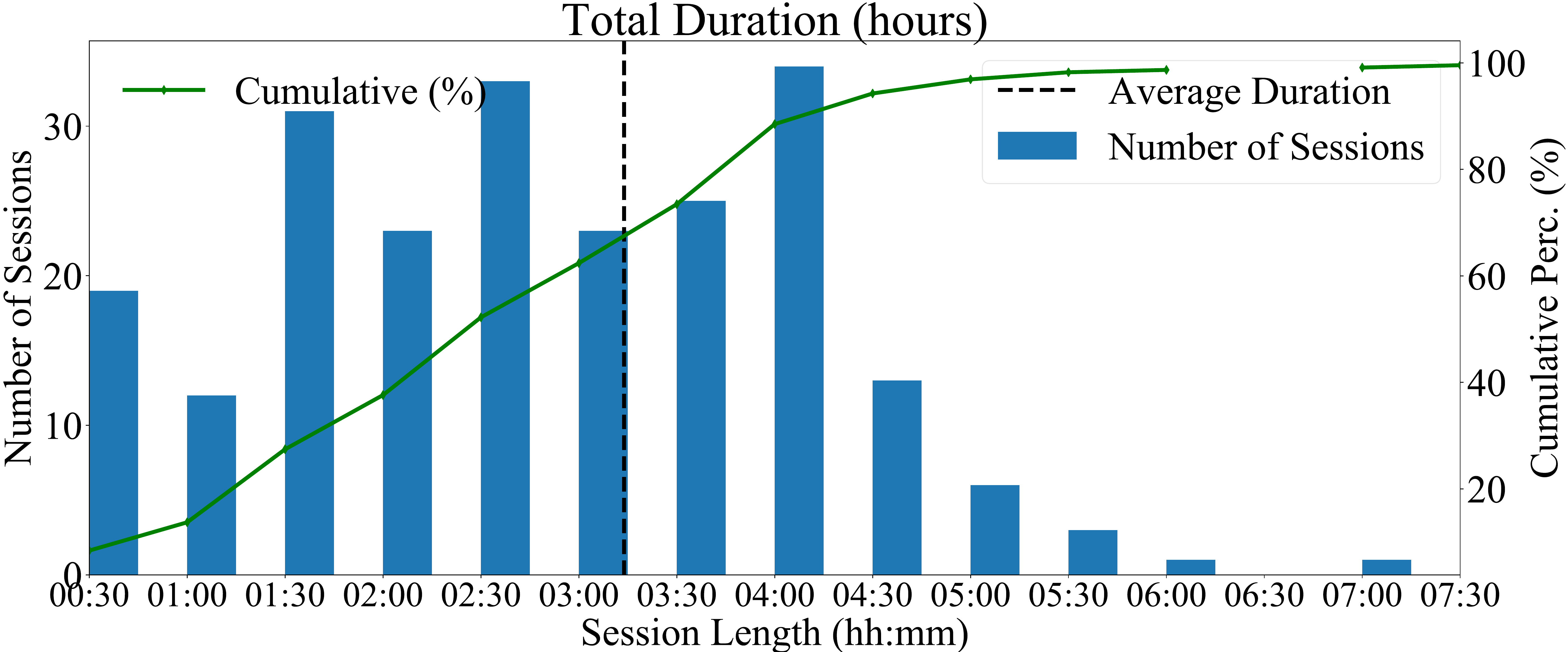}
        \caption{Histogram of total parking duration for all the charging events.}
        \label{fig:tot_duration}
    \end{subfigure}
    \hfill
    \begin{subfigure}
        \centering
        \includegraphics[width=1\columnwidth]{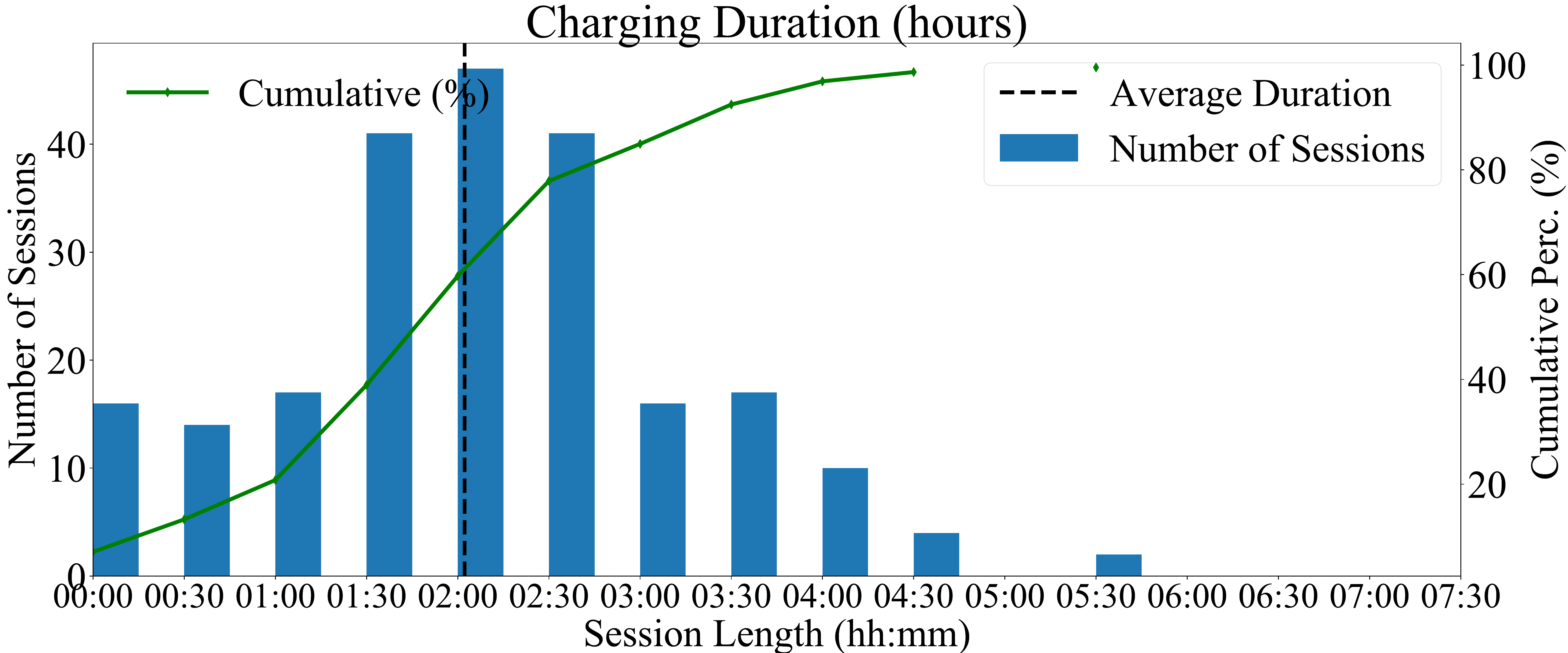}
        \caption{Histogram of charging duration for all the events.}
        \label{fig:char_duration}
    \end{subfigure}
\end{figure}

In this case study, we consider measured data from the charging facilities at Cal Poly San Luis Obispo campus, California -- a workplace charging station for faculty, staff, and students. The data was measured from January 16th, 2019 to January 23rd, 2019. There are in total 201 charging events across the week. The average parking duration is 3.25 hours (Fig.~\ref{fig:tot_duration}) and the average charging duration is less -- about 2 hours (Fig.~\ref{fig:char_duration}). %The maximum charging demand delivered is 62 kWh and the minimum is 0. 
The average charging duration to parking duration ratio is 0.62, meaning that, on average, 38\% of the time a charger is plugged-in, it does not provide service. This phenomenon results in low utilization of the charging infrastructure. 

\subsubsection{Time-of-Use Price}
The Pacific Gas \& Electric A-10, Medium General Time-of-Use Service is adopted as the time-of-use electricity tariff.

\subsubsection{Parameter Settings}
% \hl{[SB: write]}
% \begin{itemize}
%     \item parameter settings:
%     \begin{itemize}
%         \item number of charging poles at the station \hl{[TZ: 12 chargers in the real life station]}
%         \item DCM parameters
%         \item \hl{TZ: DONE} charging choices distribution
%     \end{itemize}
% \end{itemize}

% Charging station parameter
The charging station is comprised of level-2 chargers, i.e. the maximum charging power is 7.2 kW. We only consider from 7 am to 10 pm (a total of 15 hours) as the operation hours at the charging station. %We assume that all PEV drivers cannot use the charging service if they desire to park after the operation hours, which guarantees that all vehicles leave by 10 pm. 

% DCM parameter
We synthetically generated DCM parameters $\beta, \gamma,$ and $\beta_0$ so that:
\begin{itemize}
    \item \textit{charging-asap} is preferred by default;
    \item a decreasing gap in charging price between \textit{charging-flexibility} and \textit{charging-asap} options increases the probability of controlled charging;
    \item high charging prices and overstay penalty increase the probability of leaving without charging;
    \item controlled charging is preferred with high desired parking duration.
\end{itemize}
These choices are intuitive. However, they are not experimentally validated. This would require a careful design of experiments with human participants, which is outside the scope of this paper.
% Other parametres

\subsection{Simulation Results}
We run simulations across 50 days (episodes) where, in each episode, a sequence of charging events is randomly sampled from the empirical PDF of charging demand. In each charging event, each user ``randomly'' chooses a charging option according to their perceived utility, which includes charging price and the overstay penalty, computed by the pricing controller. We compare the simulation results of controlled operations, i.e., with the price controller, with the simulation results of nominal operations, i.e. without a price controller (the baseline). In particular, we specifically consider three performance metrics: (i) overstay duration, (ii) net profit, and (iii) the Quality-of-Service, which is calculated by the number of provided charging services\footnote{Note that the baseline does not consider the probability of leaving without charging as a choice option, and consequently the baseline can provide more service than the controlled case.}. %The first two factors are evaluated through \eqref{eqn:metric} and the last one is through comparing the provided number of charging events\footnote{Note that the baseline does not consider the probability of leaving without charging as a choice option, and consequently the baseline can possibly provide more service than the controlled case.}. As shown in Fig.~\ref{fig:hist}, our controller yields:

\begin{figure}
    \centering
    \includegraphics[width=1\columnwidth]{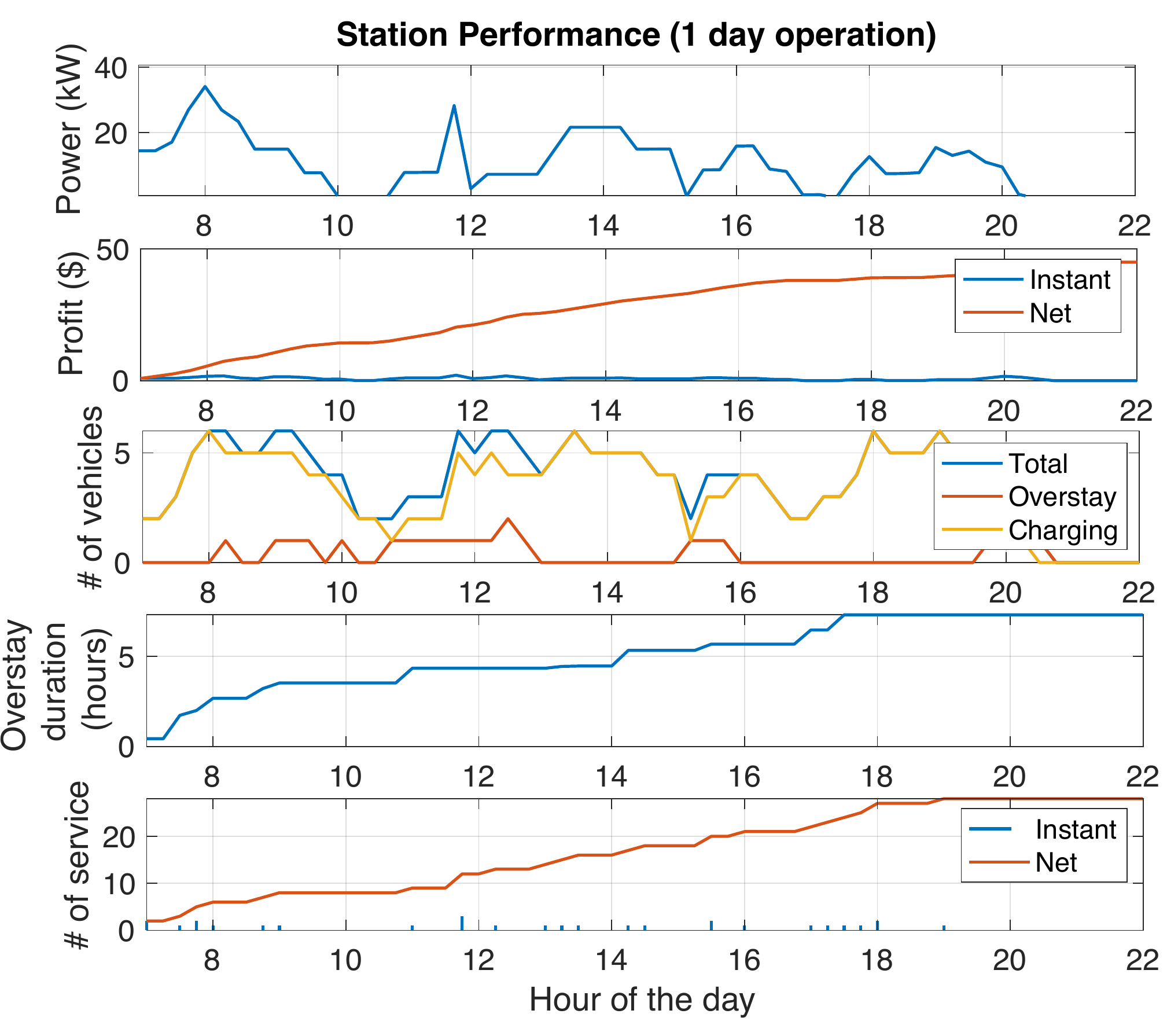}
    \caption{Simulation results of the one-day operations with the controller. Each profile indicates aggregate values over all charging poles (set to a total of 6 in the simulation).}
    \label{fig:one_day_temporal}
\end{figure}

% one day simulation
Fig.~\ref{fig:one_day_temporal} shows a single instance of the temporal profile of the charging station's total net power, profit, occupancy, cumulative overstay duration, and number of charging service, respectively, over the operating hours within a day. These plots illustrate high charging demand in the morning, where the controller is more effective at managing load, and low charging demand in the evening, where the controller provides less benefit because there is less congestion to manage. In the third plot in Fig.~\ref{fig:one_day_temporal}, one can see the overstay events primarily occur outside PG\&E's peak hour period, 13:00-18:00 (when the TOU prices are high as shown in Fig.~\ref{fig:charging_choice}). In this example, the overstaying vehicles are cleared out between 13:00-15:00, so that the charging station can accommodate more customers. This behavior occurs because the controller seeks flexible charging demand to reduce electricity consumed during PG\&E's peak pricing period, when the utility's electricity cost is high. In the bottom plot in Fig.~\ref{fig:one_day_temporal}, one can see a charging service initiates around 17:00, however, the actual charging occurs during the off-peak hours, 18:00-20:00, (top plot in Fig.~\ref{fig:one_day_temporal}). This illustrates how the optimal charging controller is managing electricity demand to minimize net costs / maximize net profit. 

\begin{figure}
    \centering
    \includegraphics[width=1\columnwidth]{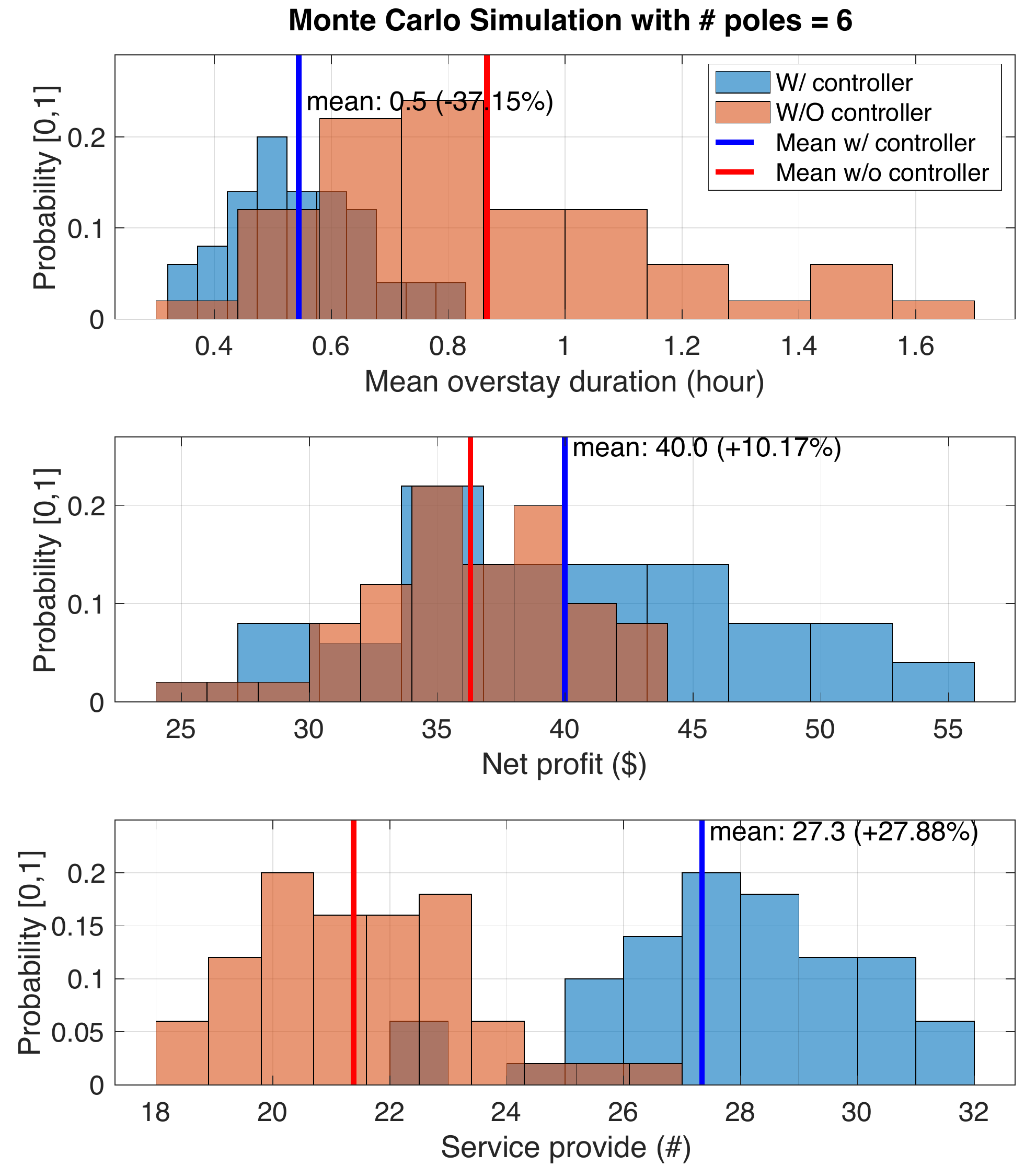}
    \caption{Monte Carlo simulation results with a total of 50 days (episodes) of operations. At each episode, a sequence of charging activities is randomly sampled from the empirical PDF generated with the Cal Poly San Luis Obispo dataset.}
    \label{fig:hist}
\end{figure}

% monte carlo
The effectiveness of the controller can be further explored through Monte Carlo simulations. Fig.~\ref{fig:hist} illustrates that the controller decreases overstay duration by 37.15\%, increases net profit by 10.17\%, and increases the number of fulfilled charging services by 27.88\%
% \begin{itemize}
%     \item 42.9\% decrease in overstay duration,
%     \item 53.24\% increase in net profit,
%     \item 58.01\% increase in number of fulfilled charging service,
% \end{itemize}
compared to the case without price control. These improvements are essentially because the controller optimally sets the overstaying penalty while mitigating the probability of losing a user. The overstay penalty encourages the users to leave sooner after their vehicle is fully charged, and therefore the station can provide more users with charging service. Consequently, the net profit increases. %Another interesting point of view is that the controller ``locally'' solves the optimization \eqref{eq:obj_compact} for each user, however, we see the ``global'' benefit of the charging station. 

That being all said, the simulations are based on a controller with perfect knowledge of the synthetically generated behavioral model, and the actual behavior will be different in practice. Therefore, instead of the numbers, we highlight the significant ``potential'' of the controller for PEV charging station operators to increase utilization.

\subsubsection{Charging Option Choices and Probabilities}
\begin{figure}
    \centering
    \includegraphics[width=1\columnwidth]{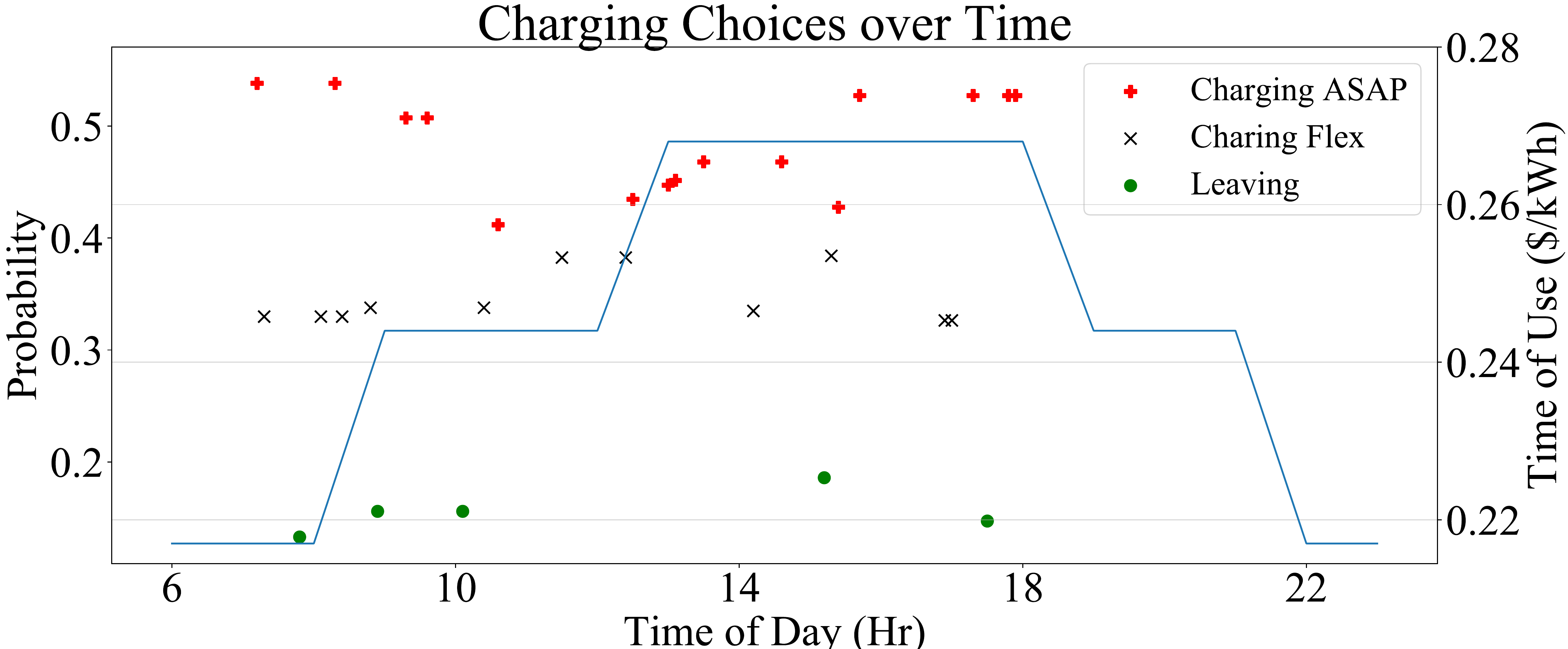}
    \caption{Charging choices and their probability over time of a day; %the optimal tariffs manipulate the probability distributions of choices over time and they are influenced by the electricity time-of-use price (TOU). 
    Blue curve indicates the evolve of TOU throughout a day.}
    \label{fig:charging_choice}
\end{figure}

% choice probability
Fig.~\ref{fig:charging_choice} shows that the users in general have higher tendency to adopt \textit{charging-asap}. However, during the utility's peak hours in the middle of the day, the controller adjusts the prices so that the preference to \textit{charging-asap} decreases while the preference to \textit{charging-flexibility} increases. To clear out overstaying vehicles during peak hours, the overstay penalty is set high and correspondingly the probability of leaving without charging increases. These results illustrate how a human behavioral model can be utilized to nudge EV charging demand for a facility operator's benefit. %However, note that the human behavior model is based on synthetically generated parameters. Future work involves human behavioral experiments to identify the parameters of a discrete choice model, for charging price control.

% As a result, human drivers' tendency to leave the station without charging increases. Our human behavior model effectively captures the dynamics of human reacting to the pricing information.

\subsection{Sensitivity Analysis}

\begin{figure}
    % \vspace{-5mm}
    \centering
    \includegraphics[scale=0.5]{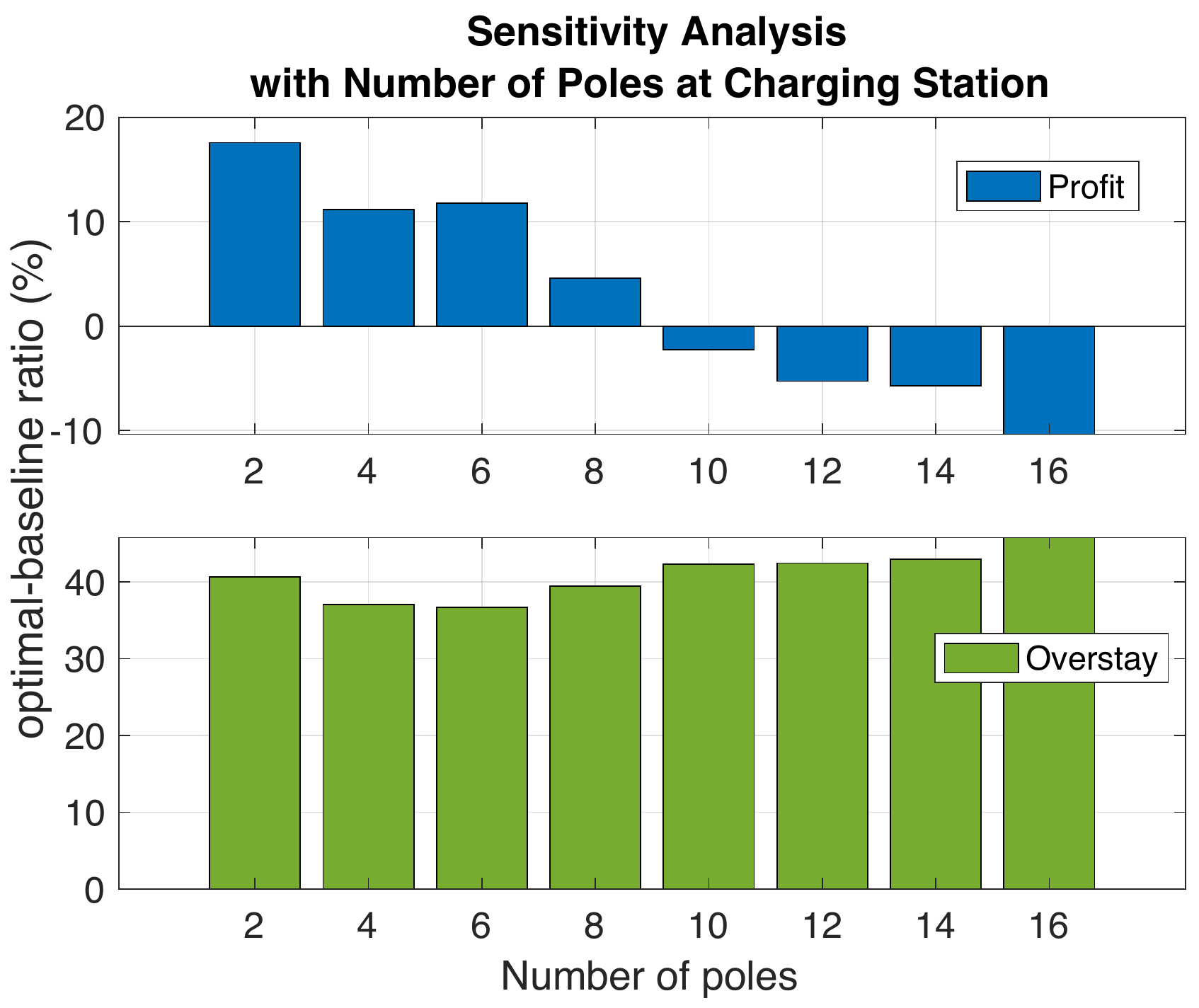}
    \caption{Sensitivity analysis by number of poles. ``Improvement" in y-axis indicates the improvement in associated property (profit or overstay) compared to the baseline (without controller), which is computed as $\left( \frac{\text{Controller Value}}{\text{Baseline Value}} - 1\right) * 100\%$.
    \label{fig:sens_barplot}}
\end{figure}

Next, we perform a sensitivity analysis by varying the number of charging poles at the station and examine the results according to the three metrics defined above. We consider the main streams of income as charging service and overstay penalty. However, with the objective to minimize overstay duration, one stream of income can be negatively impacted. As reflected in Fig. \ref{fig:sens_barplot}, when the number of charging poles is small (2-6), supply of charging infrastructure is much less than demand. In this case, the controller increases net profit by providing more charging service by clearing out more overstaying vehicles. This is also verified in our Monte Carlo simulation results in Fig. \ref{fig:hist}: there is nearly 30\% improvement in QoS, which results in more profit from providing more charging service. On the other hand, as the number of charging poles becomes large, supply of charging infrastructure exceeds demand, and this affects the overall profit gain. In this case, although it's counter-intuitive, we actually witness profit decrease. This is in particular due to the reduction of overstay duration and corresponding reduction of the profit from overstay. %In another words, profit from overstay dominates the effect of profit change when charging demands are saturated. 
Then it naturally leads to an interesting question: with given demands, what would be the optimal charging station configuration? This will be left for future work, taking the perspective of a station planner. Nevertheless, it is clear that EV charging price control yields the greatest benefits when demand exceeds supply.

\subsection{Limitations}
% \subsubsection{Limitation}
% The proposed controller does not guarantee that the resolved total cost over the entire operating hours is the system minimum. 
The proposed controller optimizes a local problem, i.e., it optimizes the costs at an individual charger when charging service is sought. We do not consider a global problem, which evaluates the total costs of the entire charging station. Nevertheless, the simulation studies demonstrate a significant improvement in the overall system-wide performance. 

In the simulation studies, We assume homogeneity among human users during the decision making process. That is, all uses have the same DCM parameters. However, each user might have different sensitivities to price information in practice. That said, it is challenging to evaluate individual behavioral models. This would involve significant efforts in tracking individual charging behaviors over a period of time to collect sufficient data to estimate the DCM parameters, as well as potential privacy issues -- a topic of on-going work. 

Apart from user homogeneity, we assume that there is no mismatch between the behavior model in the optimization problem and the actual behavior that generates choices in the simulations. The validity of this assumption depends on how accurately the DCM model represents the actual human behavior, which will also require empirical research with human subjects. 

% \begin{itemize}
%     \item We assume homogeneity among human users during decision making process.
%     \item We assume that there is no mismatch between the behavior model in the optimization problem and the actual behavior model that generates choices in the simulations. 
%     \item We do not have a guarantee that the resolved total cost over the entire operating hours is the system minimum. Our controller optimizes a low level local problem, i.e., we optimize the costs of an individual charger when charging service is sought. We do not consider a high level system problem, which evaluates the total costs of the entire charging station across time horizon.
% \end{itemize}

% \subsubsection{Future Work}
% \hl{[I suggest removing this. If you present half-baked ideas without detail and without results, then it's an invitation for reviewers to criticize. Better to just save the idea for when it's fully ready. In other words, don't be Elon Musk and promise something without evidence that it completely works.] We plan to extend and incorporate more practical setting with a waiting option. When PEV arrives at the facility and encounter chargers fully occupied situation, we provide the option to queue at the facility. This option can associate with a ``switching cost" for interchange operation, defined in} \cite{zeng2019solving}.

\section{Conclusion}\label{sec:conclusion}
This paper designs a mathematical control framework for charging station operation aimed at alleviating the overstay issue, while maximizing net profit. In the framework, we incorporate a Discrete Choice Model (DCM) from behavioral economics to quantify the probability of a user's selection, given a controllable price. The control problem is a non-convex problem, yet has a particular structure that enables reformulation into a multi-convex problem. We show that Block Coordinate Descent can effectively solve the multi-convex problem. We validate the proposed framework with a real dataset for charging demand using an agent-based simulator. Monte-Carlo simulation results indicate a significant potential for improving three performance metrics: overstay duration, net profit, and number of fulfilled charging services. A sensitivity analysis demonstrates a loss of net profit when the availability of chargers exceeds demand, but an increase in net profit when demand exceeds charger availability. %This leaves an important question to charging infrastructure planning parties: ``what is the optimal size of a station given demands?'', which remains for future work.

% Our simulation results show a variation of human decisions with respect to different factors. These will provide interesting insight to charging station operators. Our numerical results show station operator with constrained number of chargers is still able to realize three-folds benefits: overstay duration, net profit, and number of fulfilled charging services. On the contrary, there is also counter-intuitive result: station operator with larger number of chargers might result in loss of net profit while overstay issue alleviated. This leaves an important question to charging infrastructure planning parties: ``what is the optimal size of a station given demands?'', which remains for future work.

% \section*{Acknowledgement}
% The authors thank Prof. Scott Moura and Bertrand Travacca for their valuable inputs and discussions in helping to bring this project to fruition. The authors would also like to thank Prof. Ben Recht, Karl Krauth, and Lydia Liu for their insightful feedbacks at poster session.

%========================================
\bibliographystyle{IEEEtran}
\bibliography{bibliography.bib}

\begin{thebibliography}{10}
\providecommand{\url}[1]{#1}
\csname url@samestyle\endcsname
\providecommand{\newblock}{\relax}
\providecommand{\bibinfo}[2]{#2}
\providecommand{\BIBentrySTDinterwordspacing}{\spaceskip=0pt\relax}
\providecommand{\BIBentryALTinterwordstretchfactor}{4}
\providecommand{\BIBentryALTinterwordspacing}{\spaceskip=\fontdimen2\font plus
\BIBentryALTinterwordstretchfactor\fontdimen3\font minus
  \fontdimen4\font\relax}
\providecommand{\BIBforeignlanguage}[2]{{%
\expandafter\ifx\csname l@#1\endcsname\relax
\typeout{** WARNING: IEEEtran.bst: No hyphenation pattern has been}%
\typeout{** loaded for the language `#1'. Using the pattern for}%
\typeout{** the default language instead.}%
\else
\language=\csname l@#1\endcsname
\fi
#2}}
\providecommand{\BIBdecl}{\relax}
\BIBdecl

\bibitem{zeng2019solving}
T.~Zeng, H.~Zhang, and S.~Moura, ``Solving overstay in pev charging station
  planning via chance constrained optimization,'' \emph{arXiv preprint
  arXiv:1901.07110}, 2019.

\bibitem{xu2013block}
Y.~Xu and W.~Yin, ``A block coordinate descent method for regularized
  multiconvex optimization with applications to nonnegative tensor
  factorization and completion,'' \emph{SIAM Journal on imaging sciences},
  vol.~6, no.~3, pp. 1758--1789, 2013.

\bibitem{ChargePointReport}
\BIBentryALTinterwordspacing
ChargePoint, ``All roads lead to e-mobility: Insights from 10 years of electric
  vehicle charging data,'' 2017. [Online]. Available:
  \url{https://info.chargepoint.com}
\BIBentrySTDinterwordspacing

\bibitem{chinaSpeedupEV}
\BIBentryALTinterwordspacing
Y.~Liu, ``China strives to speed up development of {EV} charging stations,''
  2018. [Online]. Available:
  \url{https://www.renewableenergyworld.com/articles/2018/06/china-strives-to-speed-up-development-of-ev-charging-stations.html}
\BIBentrySTDinterwordspacing

\bibitem{charginglocations}
\BIBentryALTinterwordspacing
D.~o.~E. Alternative Fuels Data~Center, ``Electric vehicle charging station
  locations,'' 2018. [Online]. Available:
  \url{https://www.afdc.energy.gov/fuels/electricity\_locations.html}
\BIBentrySTDinterwordspacing

\bibitem{biswas2016managing}
A.~Biswas \emph{et~al.}, ``Managing overstaying electric vehicles in
  park-and-charge facilities,'' \emph{arXiv preprint arXiv:1604.05471}, 2016.

\bibitem{lambertTesla}
\BIBentryALTinterwordspacing
F.~Lambert, ``Tesla increases supercharger idle fees to decrease wait times,''
  2018. [Online]. Available:
  \url{https://electrek.co/2018/09/19/tesla-update-supercharger-idle-fees/}
\BIBentrySTDinterwordspacing

\bibitem{pentland1999modeling}
A.~Pentland and A.~Liu, ``Modeling and prediction of human behavior,''
  \emph{Neural computation}, vol.~11, no.~1, pp. 229--242, 1999.

\bibitem{arnold2013random}
L.~Arnold, \emph{Random dynamical systems}.\hskip 1em plus 0.5em minus
  0.4em\relax Springer Science \& Business Media, 2013.

\bibitem{mukhopadhyay2015wearable}
S.~C. Mukhopadhyay, ``Wearable sensors for human activity monitoring: A
  review,'' \emph{IEEE sensors journal}, vol.~15, no.~3, pp. 1321--1330, 2015.

\bibitem{bae2018acc}
S.~Bae, S.~M. Han, and S.~J. Moura, ``System analysis and optimization of
  human-actuated dynamical systems,'' in \emph{American Control Conference
  (ACC), 2018}.\hskip 1em plus 0.5em minus 0.4em\relax IEEE, 2018.

\bibitem{bae2019modeling}
S.~Bae, S.~M. Han, and S.~Moura, ``Modeling \& control of human actuated
  systems,'' \emph{IFAC-PapersOnLine}, vol.~51, no.~34, pp. 40--46, 2019.

\bibitem{bitar2016deadline}
E.~Bitar and Y.~Xu, ``Deadline differentiated pricing of deferrable electric
  loads,'' \emph{IEEE Transactions on Smart Grid}, vol.~8, no.~1, pp. 13--25,
  2016.

\bibitem{train1986qualitative}
K.~Train, \emph{Qualitative choice analysis: Theory, econometrics, and an
  application to automobile demand}.\hskip 1em plus 0.5em minus 0.4em\relax MIT
  press, 1986, vol.~10.

\bibitem{boyd2004convex}
S.~Boyd and L.~Vandenberghe, \emph{Convex optimization}.\hskip 1em plus 0.5em
  minus 0.4em\relax Cambridge university press, 2004.

\end{thebibliography}

% \begin{IEEEbiography}[{\includegraphics[width=1in,height=1.25in,clip,keepaspectratio]{}}]{Sangjae Bae}
% \end{IEEEbiography}
% \begin{IEEEbiography}[{\includegraphics[width=1in,height=1.25in,clip,keepaspectratio]{}}]{Teng Zeng}
% \end{IEEEbiography}
% \begin{IEEEbiography}[{\includegraphics[width=1in,height=1.25in,clip,keepaspectratio]{}}]{Bertrand Travacca}
% \end{IEEEbiography}
% \begin{IEEEbiography}[{\includegraphics[width=1in,height=1.25in,clip,keepaspectratio]{}}]{Scott Moura}
% \end{IEEEbiography}

\end{document}